\documentclass[aps,twocolumn,showpacs]{revtex4}
\usepackage{graphicx}   
\usepackage{latexsym}   

\newcommand{\bold}[1]{\mbox{\boldmath $#1$}} 
\newcommand{\etal}{{\em et al.}}
\newcommand{\eg}{{\em e.g.}}
\newcommand{\ie}{{\em i.e.}}
\newcommand{\MeV}{{\rm MeV}} 
\newcommand{\fm}{{\rm fm}} 
\newcommand{\eps}{\varepsilon}
\newcommand{\del}{\partial}
\newcommand{\g}{{\rm g}}
\newcommand{\bfk}{\bold{k}}
\newcommand{\bfr}{\bold{r}}
\newcommand{\bfv}{\bold{v}}

\begin{document}

\title{Spinodal decomposition during the hadronization stage at RHIC?}

\author{J\o rgen Randrup}

\affiliation{Nuclear Science Division, 
Lawrence Berkeley National Laboratory, Berkeley, California 94720, USA\\
and Department of Physics, University of Jyv\"askyl\"a, Finland}

\date{August 21, 2003}

\begin{abstract}
The expansion of strongly interacting matter formed in high-energy nuclear collisions
drives the system through the region of phase coexistence.
The present study examines the associated spinodal instability and
finds that the degree of amplification may be sufficient
to raise the prospect of using the spinodal pattern formation
as a diagnostic tool for probing the hadronization phase transition.
\end{abstract}


\maketitle

\section{Introduction}

The central goal of high-energy nuclear collisions is to explore the expected
phase transition from the familiar hadronic world to a plasma of quarks and gluons.
This phase change of strongly interacting matter
is expected to be of first order with a critical temperature $T_c$ 
in the range of $150-200~\MeV$,
although this remains to determined experimentally.

When two heavy nuclei collide at sufficiently high energies,
such as those provided by the Relativistic Heavy Ion Collider (RHIC),
the matter in the central rapidity region is formed well above the critical temperature $T_c$
in a state of rapid longitudinal expansion.
The expansion then causes the energy density $\eps$ to continually drop,
thus forcing the system through the critical phase region
towards hadronization and subsequent chemical and kinetic freezeout.
During this evolution the system maintains local equilibrium 
and one may employ fluid dynamics 
which has the advantage that
the equation of state, $p(\eps)$, enters explicitly in the equations of motion.

The starting point for the present discussion
is the observation that the occurrence of a first-order phase transition
is intimately linked with phase coexistence and associated spinodal instability.
Indeed, a first-order phase transition occurs when the appropriate thermodynamic potential
exhibits a convex anomaly \cite{convex}.
There is then an interval of energy density
throughout which the pressure has a negative derivative, $\del p/\del\eps<0$.
This anomalous behavior identifies the region of spinodal instability
where small deviations from uniformity are amplified,
as the system seeks to break its uniformity and separate into the two coexisting phases
consistent with the given $\eps$.

However, in the absence of reliable models for the transition region,
fluid dynamical studies of nuclear collisions have usually employed the (constant) pressure 
describing the phase-separated system,
thus suppressing entirely the spinodal instability
(see, for example, Refs.\ \cite{RischkeNPA595,SollfrankPRC55,HiranoPRC66}).
Employing a simple spline technique
to obtain the approximate form of the anomalous pressure curve,
the present study examines the significance of this inherent instability
and seeks to ascertain the degree to which it may generate spatial patterns
that might offer a diagnostic tool for probing the phase transition.

\section{Equation of state}

To estimate the behavior of $p(\eps)$ through the region of phase coexistence,
we employ a cubic spline function to interpolate between the hadron and plasma phases.
As is commonly done \cite{RischkeNPA595,SollfrankPRC55,HiranoPRC66},
we approximate the former phase by an ideal gas of hadrons,
while we take a bag of non-interacting quarks and gluons for the latter.

In general,
the contributions to $p$ and $\eps$ from a non-degenerate particle of mass $m$ are 
\begin{eqnarray}
p_\kappa(m) \!&=&\!  \kappa {T^4\over2\pi^2} \sum_{n=1}^\infty
{\kappa^n\over{n}^4}  \left({nm\over T}\right)^2\!\! K_2({nm\over T}) 
\asymp {\pi^2 \over 90} T^4 ,\ \\
\eps_\kappa(m) \!&=&\! \kappa {T^4\over2\pi^2} \sum_{n=1}^\infty
{\kappa^n\over{n}^4} \left[ 3\left({nm\over T}\right)^2\!\! K_2({nm\over T})\right.\\ \nonumber
&~& \hspace{5.2em} + \left.\left({nm\over T}\right)^3\!\! K_1({nm\over T})\right]
\asymp {\pi^2 \over 30} T^4 ,\
\end{eqnarray}
where $\kappa=+$ for bosons and $\kappa=-$ for fermions.

\begin{figure}[tbh]
\includegraphics[angle=0,width=3.1in]{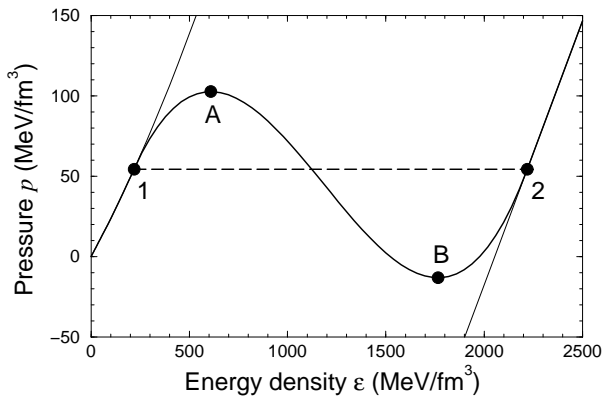}    
\caption{\label{f:eos}
The equation of state for {\em uniform} strongly interacting matter (solid)
as obtained by interpolating with a cubic spline through the coexistence region
between an ideal hadronic gas for $\eps\leq\eps_1$
and a bag with quarks and gluons for $\eps\geq\eps_2$.
The dashed line through the phase coexistence region
represents the pressure of an equilibrium {\em mixture} of the two phases.
The spinodal region extends between the two extrema of $p(\eps)$
where uniform matter is mechanically unstable, $\del p/\del\eps<0$.
}\end{figure}

The hadronic phase is approximated as an ideal gas of 14
hadronic species $i=\pi, K, \eta, \rho, \dots, N, \Lambda, \Sigma, \Delta$
with masses $m_i$ and degeneracies $\g_i$ (which include antiparticles where appropriate).
Since we are focussing on the central rapidity region of collisions at RHIC,
we shall asssume that all the chemical potentials vanish, $\mu_B=\mu_Q=\mu_S=0$.
We then have
\begin{equation}
p_{\rm had} (T) =  \sum_{i=\pi}^\Delta \g_i p_{\kappa_i}(m_i)\ ,\
\eps_{\rm had}(T)  =  \sum_{i=\pi}^\Delta \g_i \eps_{\kappa_i}(m_i)\ .
\end{equation}
The hadronic equation of state is shown in Fig.\ \ref{f:eos}.
At the critical temperature, for which we use $T_c=170~\MeV$,
the pressure is $p_c = 54~\MeV~\fm^{-3}$,
while the energy density is $\eps_1 = 220~\MeV~\fm^{-3}$.
The slope of $p(\eps)$ gives the square of the sound speed at $T_c$,
$(\del p_{\rm had}/\del\eps)_1 = v_1^2 = 0.27$.

For the plasma phase, we assume that the system can be approximated
as an ideal gas of quarks and gluons confined within a bag of suitable (negative) pressure $-B$.
The gluons as well as the $u$ and $d$ quarks are taken to be massless,
$m_g,m_q=0$,
while the $s$ quark is given a mass of $m_s=150~\MeV$
(though the results would change little if it were also zero).
The corresponding degeneracies are $\g_g=16$, $\g_q=24$, and $\g_s=12$.
Then
\begin{eqnarray}
p_{\rm qgp}(T) &=& [\g_g+\mbox{$7\over8$}\g_q]{\pi^2\over90}T^4 + \g_s p_-(m_s) -B\ ,\\
\eps_{\rm qgp}(T)  &=& [\g_g+\mbox{$7\over8$}\g_q]{\pi^2\over30}T^4 + \g_s \eps_-(m_s) +B\ .
\end{eqnarray}
Thus the pressure starts at $-B$, 
while the energy density starts at $B$.
These quantities increase more rapidly than their hadronic counterparts and
the plasma pressure overtakes the hadronic pressure at the critical temperature.
This condition determines the bag constant, $B^{1/4}=250~\MeV$.
The resulting plasma equation of state $p_{\rm qgp}(\eps)$ is shown in Fig.\ \ref{f:eos}.
The critical energy density of the plasma is 
$\eps_2 \equiv \eps_{\rm qgp}(T_c) =2220~\MeV~\fm^{-3}$,
so that the latent heat of the phase transition is 
$\Delta\eps\equiv\eps_2-\eps_1=2000~\MeV~\fm^{-3}$.
The function $p_{\rm qgp}(\eps)$ is practically linear 
(its slight non-linearity arises from $m_s$) and it has the slope
$(dp_{\rm qgp}/d\eps)_2 = v_2^2 =0.33$ at $T_c$.

Through the region of phase coexistence
we employ a cubic spline function that matches the pressure and its derivative 
(\ie\ the speed of sound) at both ends,
\begin{equation}
p(\eps)\ =\ p_c + \xi(\eps)\, [(\eps_2-\eps) v_1^2-(\eps-\eps_1) v_2^2]\, \bar{\xi}(\eps)\ ,
\end{equation}
where $\xi\equiv(\eps-\eps_1)/\Delta\eps$ 
and $\bar{\xi}\equiv (\eps_2-\eps)/\Delta\eps$.
The resulting curve is shown in Fig.\ \ref{f:eos}.
This curve rises initially at $\eps_1$ with the slope $v_1^2$ 
but it gradually bends over and turns downwards.
After crossing the mixed-phase line $p=p_c$ it starts bending upwards 
and finally joins the plasma curve at $\eps_2$ with the slope $v_2^2$.
The crossing occurs at the energy density $\eps_\times$ given by
$(v_1^2+v_2^2)\eps_\times = v_1^2 \eps_2 + v_2^2 \eps_1$.

\begin{figure}[tbh]
\includegraphics[angle=-0,width=3.1in]{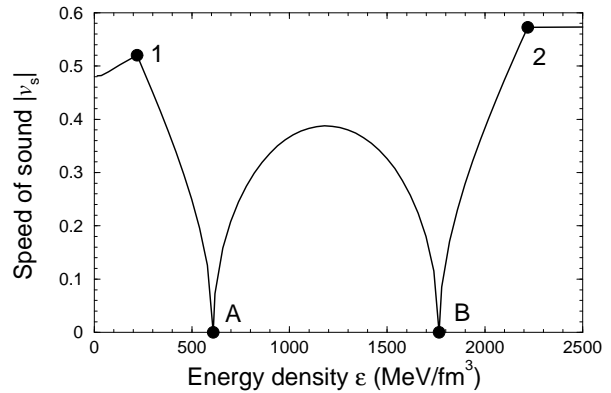}    
\caption{\label{f:vs}
The magnitude of the sound speed, $|v_s|=|\del p/\del\eps|^{1/2}$,
as a function of $\eps$ from zero to above the phase-coexistence region.
While $v_s$ is approximately constant in each of the single-phase regions,
it drops rapidly as the metastable phase-coexistence region is entered,
and becomes zero at the spinodal boundaries.
It is imaginary inside the spinodal region
where the thermodynamical potential is convex and spontaneous phase separation occurs.
In the mixed-phase scenario $v_s$ is zero 
throughout the entire phase coexistence region.
}\end{figure}

In the cubic spline approximation,
the slope $\del p/\del\eps$ is parabolic in the phase coexistence region,
\begin{equation}
 {\del p \over \del\eps}\ =\ v_s^2\
=\ \left[ 1-4\left({\eps-\tilde{\eps} \over \eps_B-\eps_A}\right)^2 \right]\ \tilde{v}_s^2\ .
\end{equation}
The inflection point in $p(\eps)$ is at $\tilde{\eps}=(\eps_1+\eps_\times+\eps_2)/3$ 
and its extrema (located at $\eps_A$ and $\eps_B$) are separated by
\begin{equation}
\eps_B-\eps_A = \mbox{$\sqrt{2}\over3$}\left[
(\eps_2-\eps_1)^2 + (\eps_\times-\eps_1)^2 + (\eps_2-\eps_\times)^2 \right]^{1\over2} .
\end{equation}
The steepest slope of $p(\eps)$ occurs at $\tilde{\eps}$ and is given by
\begin{equation}
\left({\del p\over\del\eps}\right)_{\tilde{\eps}} = \tilde{v}_s^2 =  -\mbox{$3\over4$}
\left[ v_1^2+v_2^2 \right] \left({\eps_B-\eps_A\over\Delta\eps}\right)^2 \approx\-0.150\ .
\end{equation}
This value is rather robust against changes in $T_c$ and $m_s$.
The $\eps$ dependence of the sound speed is shown in Fig.\ \ref{f:vs}.

\section{Onset of spinodal decomposition}

The system is regarded as an ideal fluid with no conserved charges.
The energy-momentum tensor $T^{\mu\nu}$ is then of ideal gas form \cite{LL},
\begin{equation}\label{Tmunu}
T^{\mu\nu}\ =\ (\eps+p) u^\mu u^\nu - p g^{\mu\nu}\ ,
\end{equation}
where $u^\mu=\gamma(1,\bfv)$ is the four-velocity of the energy flow,
and  local conservation of four-momentum yields the basic equations of motion,
$\del_\mu T^{\mu\nu}=0$, \ie\
\begin{eqnarray}\label{eq0}
\del_t [\eps+p)\gamma^2-p] &=& \bold{\nabla}\cdot [(\eps+p)\gamma^2 \bfv]\ ,\\ \label{eqx}
\del_t [(\eps+p)\gamma^2 \bfv]\,\,\,\, &=& \bold{\nabla}[(\eps+p)\gamma^2 v^2 +p]\ .
\end{eqnarray}
These coupled equations are closed by the specification of the equation of state $p(\eps)$.

We now consider a small perturbations on a uniform static system,
\begin{equation}\label{eps}
\eps(t,\bfr)\ =\ \eps_0 + \delta\eps(t,\bfr)\ ,\ \delta\eps(t,\bfr)\ll\eps_0\ .
\end{equation}
Since the flow speed $v$ is of the order of $\delta\eps/\eps_0$,
it may be regarded as small, $v\ll1$,
and we may ignore terms of order $v^2$ and thus use $\gamma=1$.
Insertion of $\eps(t,\bfr)$ from (\ref{eps}) into eqs.\ (\ref{eq0}-\ref{eqx}) then yields 
\begin{eqnarray}\label{nu0}
0\! &\approx&\! \del_t \eps - \bold{\nabla}\!\cdot\![(\eps+p) \bfv]
\approx \del_t\delta\eps - (\eps_0+p_0) \bold{\nabla}\!\cdot\bfv,\\ \label{nux}
\bold{0}\! &\approx&\! \del_t [(\eps+p) \bfv] - \bold{\nabla}p 
\approx (\eps_0+p_0)\del_t \bfv - v_s^2\, \bold{\nabla}\delta\eps,\,\,\
\end{eqnarray}
where we have used 
$\bold{\nabla}p(\eps(\bfr))=(\del p_0/\del\eps_0)\bold{\nabla}\eps=v_s^2\bold{\nabla}\eps$,
with $p_0\equiv p(\eps_0)$ being the pressure in the unperturbed uniform system.
The speed of sound was displayed in Fig.\ \ref{f:vs}.
Taking the gradient of (\ref{nux}) and using the derived relation (\ref{nu0})
between $\delta\eps$ and $v$ then yields
the familiar equation for the propagation of sound waves,
\begin{equation}
\del_t^2\delta\eps(t,\bfr)\ =\  {\del p_0\over \del\eps_0}\ \nabla^2 \delta\eps(t,\bfr)\ .
\end{equation}
For harmonic perturbations, $\delta\eps\sim\exp(i\bfk\!\cdot\!\bfr-i\omega t)$, 
it yields a linear relation between the wave number and the frequency,
$\omega_k = v_s k$.
Thus, if a physical system is brought into the spinodal region 
where $\del p_0/\del\eps_0<0$
(such a quenching can be accomplished by rapid expansion and/or cooling),
the frequency is imaginary and small deviations from uniformity evolve exponentially.

The linear dispersion relation
becomes unrealistic once the wave length of the perturbation, $\lambda_k=2\pi/k$,
has shrunk to the size of the range of the interaction between the constituents of the fluid.
At such small scales, the physical justification for employing fluid dynamics is questionable.
However, a physically meaningful treatment can still be made within the framework 
of fluid dynamics by employing a suitable smoothing, as we shall now explain.

The equation of state $p(\eps)$ gives the pressure in a spatially uniform system
as a function of its energy density.
When the energy density varies sufficiently gently with position, 
one may obtain the corresponding local pressure as $p(\bfr)\approx p(\eps(\bfr))$,
\ie\ one may calculate $p$ as if the energy density had everywhere the same value as at $\bfr$.
This local density approximation becomes increasingly inaccurate as the scale of the
spatial variations in $\eps(\bfr)$ is decreased,
because the finite range of the forces reduces the response to fine ripples.
A simple way to account approximately for the effect of the interaction range
is to average the pressure over the interaction volume.
While this procedure is in fact exact in certain cases
(\eg\ semiclassical one-body models of nuclear systems \cite{fold}),
it is adopted here merely as a simple and transparent approximation
which needs ultimately to be justified/improved on the basis on microscopic models
of the system.

\begin{figure}[tbh]
\includegraphics[angle=-0,width=3.1in]{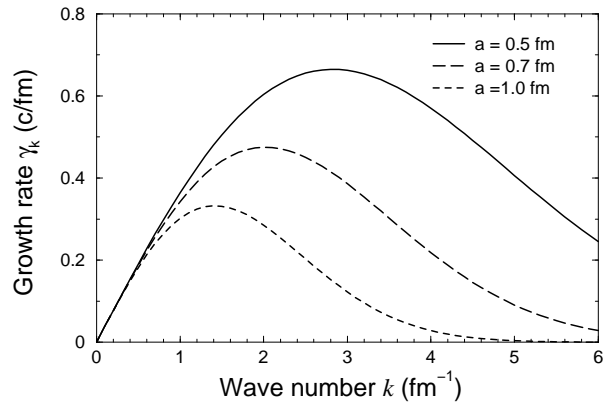}    
\caption{\label{f:gammak}
The spinodal dispersion relation (\ref{disp})
at the energy density $\tilde{\eps}$ where the pressure decreases most rapidly.
Results are shown for three values of the range $a$.
}\end{figure}

Thus we assume that the local pressure $p(\bfr)$ can be obtained by a convolution
with a (normalized) kernel $g(r)$,
\begin{equation}
p(\bfr_1)\ \approx \int d^3\bfr_2\ g(|\bfr_1-\bfr_2|)\ p(\eps(\bfr_2))\ .
\end{equation}
When the perturbation of the energy density is harmonic, $\delta\eps(\bfr) \sim \exp(i\bfk\cdot\bfr)$,
the gradient of the local pressure is then proportional to the gradient of the energy density,
\begin{equation}
\bold{\nabla} p(\bfr)\ \approx\  
{\del p_0 \over \del\eps_0} \, g_k\, \bold{\nabla} \eps(\bfr)\ 
=\ i\bfk g_k\, {\del p_0 \over \del\eps_0} \, \delta\eps(\bfr)\ ,
\end{equation}
where $g_k$ is the Fourier transform of $g(r)$.
Hence the refinement amounts to simply modulating the local-density term
$\del p_0/\del\eps_0$ in the dispersion relation by $g_k$, yielding
\begin{equation}\label{disp}
\omega_k^2 = {\del p_0 \over \del\eps_0}\ g_k\ k^2:\
\gamma_k = |v_s(\eps_0)|\, k \sqrt{g_k}\ .
\end{equation}
As a result of this convenient factorized form,
the largest collective frequency occurs at the same wave number $k_0$ for any energy density,
thus enhancing the formation of a distinctive spinodal pattern of a characteristic scale.

The spatial scale of the pattern is determined by the range of $g(r)$,
which one would expect to be in the range of $0.5-1.0~\fm$,
with the precise value depending on the specific microscopic treatment.
For simplicity, we employ here a gaussian form factor,
$g(r)\sim \exp(-r^2/2a^2)$, so we have $g_k=\exp(-a^2k^2/2)$.
The growth rates $\gamma_k$ for the spinodal modes
can then be written on a simple form,
\begin{equation}\label{gammak}
\gamma_k(\eps) = \left(-{\del p \over \del\eps}\right)_{\tilde{\eps}}^{1\over2}
\left[1-4\left({\eps-\tilde{\eps} \over \eps_B\!-\!\eps_A}\right)^2\right]^{1\over2}
k\,{\rm e}^{-{1\over4}a^2k^2} .
\end{equation}
This dispersion relation is illustrated in Fig.\ \ref{f:gammak}.

\begin{figure}[t]
\includegraphics[angle=-0,width=3.1in]{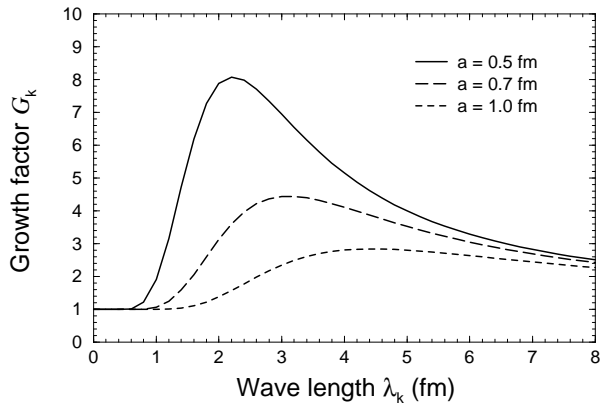}    
\caption{\label{f:Gk}
The factor $G_k=\exp(\Gamma_k)$
by which the amplitude of undulations of wave length $\lambda_k=2\pi/k$
grows during the expansion through the spinodal region.
}\end{figure}

As the system expands and the energy density traverses the spinodal region,
the growth rate $\gamma_k(\eps(t))$ of a given mode increases from zero at $\eps_B$,
exhibits a maximim at $\tilde{\eps}$, and then reverts to zero at $\eps_A$.
The total growth factor is then approximately equal to $G_k=\exp(\Gamma_k)$
\cite{HeiselbergPRL61},
with the amplification coefficient $\Gamma _k$ being
\begin{equation}
\Gamma_k\ \equiv\ \int_{t_B}^{t_A} \gamma_k(\eps(t))\, dt\
\approx\ \gamma_k(\tilde{\eps})\, \Delta t_{\rm eff}\ .
\end{equation}
The effective time during which spinodal instability is encountered
is $\Delta t_{\rm eff} \approx (\pi/4)(t_B-t_A)$ in the cubic spline approximation.
For a quantitative estimate we use the value $t_B-t_A\approx4\,\fm/c$
gleaned from Ref.\ \cite{HiranoPRC66}.
The resulting growth factors $G_k$ are displayed in Fig.\ \ref{f:Gk}.

After rapidly reaching a maximum, $G(\lambda)$ subsides relatively slowly,
thus allowing some degree of amplification of longer wave lengths.
If $a=0.5\,\fm$, the maximum growth factor of $G_{\rm max}\approx8$ 
is obtained  for $\lambda\approx2\,\fm$,
while $a=0.7\,\fm$ yields $G_{\rm max}\approx4$ for $\lambda\approx3\,\fm$.
The corresponding component of the density-density correlation function 
is proportional to the square of that,
$\langle\delta\eps(\bfr_1)\,\exp(i\bfr_{12}\cdot\bfk)\,\delta\eps(\bfr_2)\rangle \sim G_k^2$,
and can thus reach quite appreciable magnitudes.
(Without the finite-range modulation factor $g_k$
there would be even larger amplification but no characteristic scale would emerge.)

\section{Discussion}

Spindoal decomposition is intimately linked to the occurrence of a first-order phase transition
and is important in many areas of physics as a mechansim for pattern formation.
In nuclear physics it plays a key role for nuclear multifragmentation 
where it gives rise to highly non-statistical fragment size distributions
with a preference for equal masses \cite{HeiselbergPRL61,BorderiePRL86}.
The present study is focussed on high-energy nuclear collisions
and we have investigated the importance of spinodal decomposition 
while the expanding matter passes through the phase coexistence region.
Working within the framework of relativistic fluid dynamics
and basing our analysis on commonly employed assumptions about the equation of state
(supplemented with a simple spline procedure),
we have found that a significant degree of amplification may occur
while the matter passes through the mechanically unstable region of phase coexistence.

Such an onset of spinodal decomposition may have interesting dynamical consequences,
most notably the development of characteristic patterns in the energy density,
as the most rapidly amplified wavelengths grow predominant.
The spatial scale of any such emerging spinodal pattern 
could in principle be probed by HBT interferometry.
It is also conceivable that the azimuthal multipolarity of the favored modes
would be reflected in the flow coefficients $v_m$
(of which $v_2$ quantifying elliptic flow is the best explored \cite{v2}).
Investigation of these prospects would clearly be interesting.

If the appearance of spinodal patterns could in fact be examined experimentally,
one could obtain fairly direct information on the equation of state in the critical region, 
since the relative prominence of various scales would reflect 
the degree of spinodal instability encountered by the corresponding wave lengths.
Conversely, if the absence of such pattens can be established 
it may be concluded that no spinodal region has been encountered and,
consequently, the phase transition cannot be of first order.

In view of the potential utility of spinodal decomposition as a diagnostic tool,
it would seem worthwhile to pursue this issue with more realistic dynamical treatments
that take better account of the spatio-temporal features of the system
(including both the rapid longitudinal expansion and the finite transverse extent
which may affect the present simple estimates).
In particular,
the quantitative importance of spinodal decomposition
could be elucidated with existing fluid dynamical models that have been modified
to accomodate an equation of state containing the convexity anomaly
which is a characteristic feature of first-order phase transitions.

\section*{Acknowledgements}

I wish to acknowledge helpful discussions with 
K.J.\ Eskola, T.\ Hirano, P.\ Huovinen, P.V.\ Ruuskanen, B.\ Schlei, and R.\ Vogt.
This work was supported by the Office of Energy Research,
Office of High Energy and Nuclear Physics,
Nuclear Physics Division of the U.S. Department of Energy
under Contract No.\ DE-AC03-76SF00098.



                        \end{document}